\documentstyle[epsfig]{article}
= -2truecm
\begin{document}
\begin{center}

{\large\bf Exact diagonalization for spin-${1/2}$ chains and the
first order quantum\\ phase transitions of the XXX chain in a
uniform transverse field} \vskip .6cm {\normalsize Feng
Pan,$^{a,~b}$ ~Xin Guan,$^{a}$~ Nan Ma,$^{a}$ Wen-Juan Han,$^{a}$
~and J. P. Draayer$^{b}$} \vskip .2cm {\small $^{a}$Department of
Physics, Liaoning Normal University, Dalian 116029, P. R.
China\vskip .1cm $^{b}$Department of Physics and Astronomy,
Louisiana State University, Baton Rouge, LA 70803-4001, USA}
\end{center}

\begin{abstract} A simple Mathematica code based on the differential
realization of hard-core boson operators for finding exact solutions
of the periodic-$N$ spin-${1/2}$ systems with or beyond nearest
neighbor interactions is proposed, which can easily be used to study
general spin-${1/2}$ interaction systems. As an example, The code is
applied to study XXX spin-${1/2}$ chain with nearest neighbor
interaction in a uniform transverse field. It shows that there are
$[N/2]$ level-crossing points in the ground state, where $N$ is the
periodic number of the system and $[x]$ stands for the integer part
of $x$, when the interaction strength and magnitude of the magnetic
field satisfy certain conditions. The quantum phase transitional
behavior in the ground state of the system in the thermodynamic
limit is also studied.
\end{abstract}

As is well-known, the finite periodic spin-${1/2}$ chain with
nearest neighbor interaction in a uniform transverse field is
exactly solvable by using either Bethe ansatz or transfer matrix
techniques.$^{[1-6]}$ Similar spin chain models have been attracted
a lot of attention recently due to the fact that they may be
potentially helpful in quantum information processing$^{[7-9]}$ and
realizable by using quantum dots, optical lattice, or spin
interaction systems, {\it etc}.$^{[10-12]}$ Quantum phase
transitions (QPTs) and entanglement in these systems are of great
interest because there are intimate links between the QPTs and
entanglement.$^{[9, 14-16]}$ Though numerical Bethe ansatz solution
to the problem is possible and helpful in the large $N$ limit, it is
too complicated and difficult to be compiled into a practical
algorithm for large but finite $N$ cases. More importantly, there is
still in need of a simple approach to exact solutions of spin
systems beyond nearest neighbor interactions. In this Letter, we
report our exact diagonalization algorithm for spin-${1/2}$ systems
written in Mathematica by using differential realization of the
hard-core boson operators. The simple code can easily be used to
study general one-dimensional spin-${1/2}$ interaction systems, such
as XY or XYZ spin-${1/2}$ chains. As an example, The code is applied
to study XXX spin-${1/2}$ chain with nearest neighbor interaction in
a uniform transverse field, which shows that there are a series of
level-crossing points when the interaction strength and magnitude of
the magnetic field satisfy certain conditions similar to the
situation in the XX spin chain studied in [17].  The entanglement
measure$^{[18,19]}$ defined in terms of von Neumann entropy of
one-body reduced density matrix is used to measure the
multi-particle entanglement and reveal the QPTs in the system.

By using the hard-core boson mapping: $S^{+}_{i}=S^{x}_{i}+\imath
S^{y}_{i}\rightarrow b^{\dagger}_{i}$, $S^{-}_{i}=S^{x}_{i}-\imath
S^{y}_{i}\rightarrow b_{i}$, and $S^{0}_{i}=S^{z}_{i}\rightarrow
b^{\dagger}_{i}b_{i}-{1\over{2}}$, where $S^{\mu}_{i}$ ($\mu=x,y,z$)
are spin operators satisfying the SU(2) commutation relations, the
periodic condition $S^{\mu}_{i+N}=S^{\mu}_{i}$ is assumed, $b_{i}$
and $b^{\dagger}_{i}$ satisfy $[b_{i},
b_{j}^{\dagger}]=\delta_{ij}(1-2b^{\dagger}_{j}b_{j})$,
$[b^{\dagger}_{i},b^{\dagger}_{j}]=[b_{i},b_{j}]=0$, and
$(b_{i})^{2}=(b^{\dagger}_{i})^2=0$, the Hamiltonian of the XXX
spin-${1\over{2}}$ chain with nearest neighbor interaction in a
uniform transverse field, for example, can then be written as

$$H_{\rm XXX}={J}\sum^{N}_{i=1}\left({1\over{2}}(b^{\dagger}_{i}b_{i+1}+
b^{\dagger}_{i+1}b_{i})+(b^{\dagger}_{i}b_{i}-{1\over{2}})
(b^{\dagger}_{i+1}b_{i+1}-{1\over{2}})\right)+h\sum^{N}_{i=1}\left(
b^{\dagger}_{i}b_{i}-{1\over{2}}\right).\eqno(1)$$ where $J>0 ~(<0)$
corresponds to the anti-ferromagnetic (ferromagnetic) case, and $h$
is a uniform transverse field. Then, by using the differential
realizations for the boson operators with
$b^{\dagger}_{i}\rightarrow x_{i}$, $b_{i}\rightarrow\partial_{i}$,
(1) can be rewritten as

$$H_{\rm XXX}={\cal P}\left({J}\sum^{N}_{i=1}\left({1\over{2}}(x_{i}\partial_{i+1}+
x_{i+1}\partial_{i})+(x_{i}\partial_{i}-{1\over{2}})
(x_{i+1}\partial_{i+1}-{1\over{2}})\right)+h\sum^{N}_{i=1}\left(
x_{i}\partial_{i}-{1\over{2}}\right)\right){\cal P},\eqno(2)$$ where
${\cal P}$ is an operation to project a state with $x^{q}_{i}=0$ for
$q\geq 2$ ( $i=1,2,\cdots,n$) due to the hard-core restriction. One
can easily verify that the differential realization with such
restriction is consistent to the commutation relations of the
hard-core boson operators.

Because the total number of bosons,
$\hat{k}=\sum_{i=1}^{N}b^{+}_{i}b_{i}$ is conserved, $k$-`particle'
wavefunction of (2) can be expressed in terms of $k$-th order
homogenous polynomials of $\{x_{i}\}$ with

$$F^{(\zeta)}_{k}(x_{1},\cdots,x_{N})=\sum_{1\leq i_{1}<
i_{2}<\cdots< i_{k}\leq N} C^{(\zeta)}_{i_{1} i_{2}\cdots
i_{k}}x_{i_{1}}x_{i_{2}}\cdots x_{i_{k}},\eqno(3)$$ where
$C^{(\zeta)}_{i_{1} i_{2}\cdots i_{k}}$ is the expansion
coefficient, and $\zeta$ is used to label different eigenstate with
the same $k$. Using (2) and (3), one can establish the
eigen-equation

$$H_{\rm XXX}F^{(\zeta)}_{k}(x_{1},\cdots,x_{N})=E^{(\zeta)}_{k}
F^{(\zeta)}_{k}(x_{1},\cdots,x_{N})\eqno(4)$$ which is a second
order linear partial differential equation and can easily be solved
with a Mathematica code.$^{[20]}$ It should be stated that the first
projection ${\cal P}$ at the end of (2) becomes an identical
operation since there is no $x_{i}^{q}$ with $q\geq 2$ occurring in
(3), while the final projection ${\cal P}$ should be considered in
the code, which can simply be realized by setting $x_{i}=0~\forall
~i$ after the matrix elements of the Hamiltonian being constructed.
Though only an example with $N=8$ and $k=3$ for the XXX spin chain
with nearest neighbor interaction Hamiltonian $H_{\rm XXX}/J$ with
$h=0$ is shown in [20], it is obvious that the procedure shown in
[20] can easily be extended to more general cases, such as XY or XYZ
spin-${1\over{2}}$ chain models with or beyond nearest neighbor
interaction. It can be seen from [20] that we first construct the
eigen-equation of XXX model Hamiltonian in the $x$-representation.
Then, we can obtain the energy sub-matrix for any $k$, which can be
output to other code for diagonalization.  Hence, the original
$2^{N}$ dimensional energy matrix is reduced to $N!/(N-k)!k!$
dimensional submatrices.  Once the eigenenergy $E^{(\zeta)}_{k}/J$
and the corresponding eigenvector $\{C^{(\zeta)}_{i_{1}i_{2}\cdots
i_{k}}\}$ are known after diagonalization, the final wavefunction
can be expressed as

$$\vert
k;\zeta\rangle=F^{(\zeta)}_{k}(b^{\dagger}_{1},\cdots,b^{\dagger}_{N})\vert
0\rangle,\eqno(5)$$  where $\vert 0\rangle$ is the boson vacuum and
thus the SU(2) lowest weight state with $S_{i}^{-}\vert
0\rangle=0~\forall~i$.

 As an example of application of the code, in the following,
we study quantum phase transitional behavior of the finite periodic
XXX spin-${1\over{2}}$ chain with nearest neighbor interaction in a
uniform transverse field. One can verify that there is no quantum
phase transition for the ferromagnetic case with $J<0$, in which the
ground state of the system with $J<0$ keeps unchanged in the
variation of the magnitude of the magnetic field. Quantum phase
transition occurs only in the anti-ferromagnetic cases with $J>0$,
which will be considered in the following. In order to investigate
QPT behavior of the system for $J>0$, we set $J=1-t$ and $h=t$ with
$0\leq t\leq 1$. It is clear that the ground state of the system is
in the ferromagnetic  (unentangled) phase when $t=1$ and in the
anti-ferromagnetic long-range order (entangled) phase when $t=0$.
Therefore, $t$ serves as the control parameter of the system. In the
XXX case, in addition to $S_{0}=k-N/2$, the total spin of the system
$S$ is also a good quantum number. Therefore, the wavefunction (5)
can further be written as $\vert S_{0}=k-N/2;S,\xi\rangle$, where
the additional quantum number $\xi$ is used to label different
eigenstate with the same $S$ and $S_{0}$. Though one can only obtain
an eigenstate with fixed $S_{0}$ from the code, one may get
information about the total spin $S$ by acting on the total spin
lowering operator $S^{-}=\sum_{i=1}^{N}S^{-}_{i}$ to the state. For
example, the state $\vert S_{0}=-N/2; S=N/2,\xi\rangle$ must satisfy
$S^{-}\vert S_{0}=-N/2; S=N/2,\xi\rangle=0$, while $\vert
S_{0}=1-N/2; S=N/2-1,\xi\rangle$ must satisfy $S^{-}\vert
S_{0}=1-N/2; S=N/2-1,\xi\rangle=0$, and so on, which enables us to
find the corresponding quantum number $S$ for each eigenstate. For
$N$ odd cases, the eigenstates with $S_{0}=k-N/2$ and $S=N/2-k$ for
$k\neq 0$ are doubly degenerate. In such cases, the expansion
coefficients $\vec{C}(\xi)=\{C^{(\xi)}_{i_{1} i_{2}\cdots i_{k}}\}$
with $\xi=1$ and $\xi=2$ obtained from the code are not orthogonal
with each other. In such cases, we use the Gram-Schmidt
orthogonalization procedure to set $\vec{C'}(\xi=1)=
\vec{C}(\xi=1)-\vec{C}(\xi=1)\cdot \vec{C}(\xi=2) \vec{C}(\xi=2)$
and keep $\vec{C}(\xi=2)$ unchanged after normalization.

It is well known that the ground state of the anti-ferromagnetic XXX
spin chain is never degenerate with $S=0$ for $N$ even and four-fold
degenerate with degeneracy equal to $2(2S+1)$ and $S=1/2$ for $N$
odd, which all correspond to $t=0$. We have verified that the ground
state energy of the system is related to the following set of
eigen-energies: $E^{S=N/2-k}_{S_{0}=-S,~\min}(t)\equiv
E^{k}_{\min}(t)$ for $k=0,1,\cdots,[N/2]$, where $[x]$ stands for
the integer part of $x$. It should be stated that the ground state
energy at $t=1$ corresponds to $E^{k=0}_{\min}(t)$, while that at
$t=0$ corresponds to $E^{k=[N/2]}_{\min}(t)$. Hence, it is clear
that there are also $[N/2]+1$ different ground states which are
mutually orthogonal with the corresponding ground state energy
$E^{k=0}_{\min}(t)$, $E^{k=1}_{\min}(t)$, $\cdots$,
$E^{k=[N/2]}_{\min}(t)$ when the control parameter $t$ changes from
$1$ to $0$ similar to the situation of XX spin-${1\over{2}}$ chain
reported in [17]. Obviously, the quantum phase transitions occurring
in such cases are of the first order.$^{[17]}$  It can be verified
by the code that all levels with eigenenergy $E^{k}_{\min}(t)$ for
$k=1,2,\cdots,[N/2]$ are not degenerate for $N$ even and $0\leq
t\leq 1$, while they are all two-fold degenerate for $N$ odd and
$0<t<1$. The degeneracy of the ground state for $N$ odd at $t=0$ is
$2(2S+1)=4$, while the ground state for $N$ odd at $t=1$ is a
singlet with $S=N/2$ and $S_{0}=-N/2$.

\vskip .3cm \noindent{\bf Table 1.}~{$[N/2]$ level-crossing points
for $2\leq N\leq 12$.}\\
\begin{tabular*}{\textwidth}{ccccccccccccc}
\hline \hline $N$~&$t^{(1)}_{\rm c}$& $t^{(2)}_{\rm c}$~
&$t^{(3)}_{\rm c}$ &$t^{(4)}_{\rm c}$& $t^{(5)}_{\rm c}$~
&$t^{(6)}_{\rm c}$\\
\hline\\
2&0.666666\\
3&0.600000\\
4&0.500000&0.666666 \\
5&0.566915&0.644004\\
6&0.499123&0.566401&0.666666 \\
7&0.511933&0.623396&0.655288\\
8&0.343259&0.570166&0.643104&0.666666\\
9&0.462701&0.591992&0.642284&0.659828\\
10&0.297378&0.527473&0.614872&0.652704&0.666666\\
11&0.420934&0.559842&0.621991&0.650981&0.662104\\
12&0.262455&0.490059&0.58657&0.634069&0.657415&0.666666\\
\hline \hline\\
\end{tabular*}
\vskip .3cm The first order phase transition in the system occurs
due to the ground state energy level-crossing of $E^{i}_{\min}(t)$
with $E^{i+1}_{\min}(t)$ for $i=0,1,\cdots,[N/2]-1$ with the
corresponding critical point $t^{([N/2]-i)}_{\rm c}$, which is the
root of the simple linear equation $E^{i}_{\min}(t^{([N/2]-i)}_{\rm
c})=E^{i+1}_{\min}(t^{([N/2]-i)}_{\rm c})$ for $i=0,1,2,\cdots,
[N/2]-1$. There are $[N/2]$ such level-crossing points indicating
that there are $[N/2]+1$ different ground states within the control
parameter range $0\leq t\leq 1$. Fig. 1 clearly shows the ground
state level-crossings in the entire control parameter range for
$N=2,4,~5,~6,~8$, and $12$ cases. It is obvious that there are
$[N/2]$ level-crossing points dividing the ground state into
$[N/2]+1$ different parts, of which each is within a specific $t$
range when $N$ is a finite number. With $N$ increasing, however,
these specific ranges become smaller and smaller, and finally tend
to infinitesimal, thus the ground state level becomes a continuous
phase before crossing to $E^{0}_{\min}$ level. Therefore, there will
be only one obvious critical point when $N\rightarrow\infty$. One
can verify that the critical point $t_{\rm c}^{([N/2])}=2/3$ is
$N$-independent for $N$ even, while it will tend to $2/3$ for $N$
odd when $N\rightarrow\infty$. Nevertheless, other level-crossing
point $t^{(i)}_{\rm c}$ values  are $N$-dependent, of which some
examples are listed in Table 1.

\begin{center}
\begin{figure}[lh]
\center{\epsfig{file=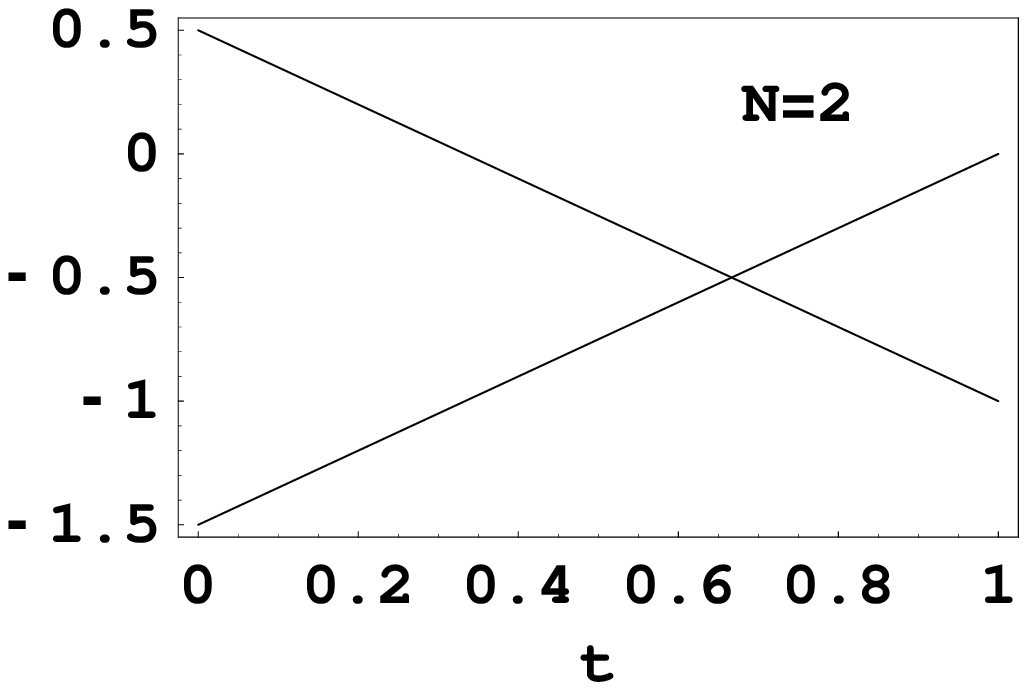,width=5cm}
\epsfig{file=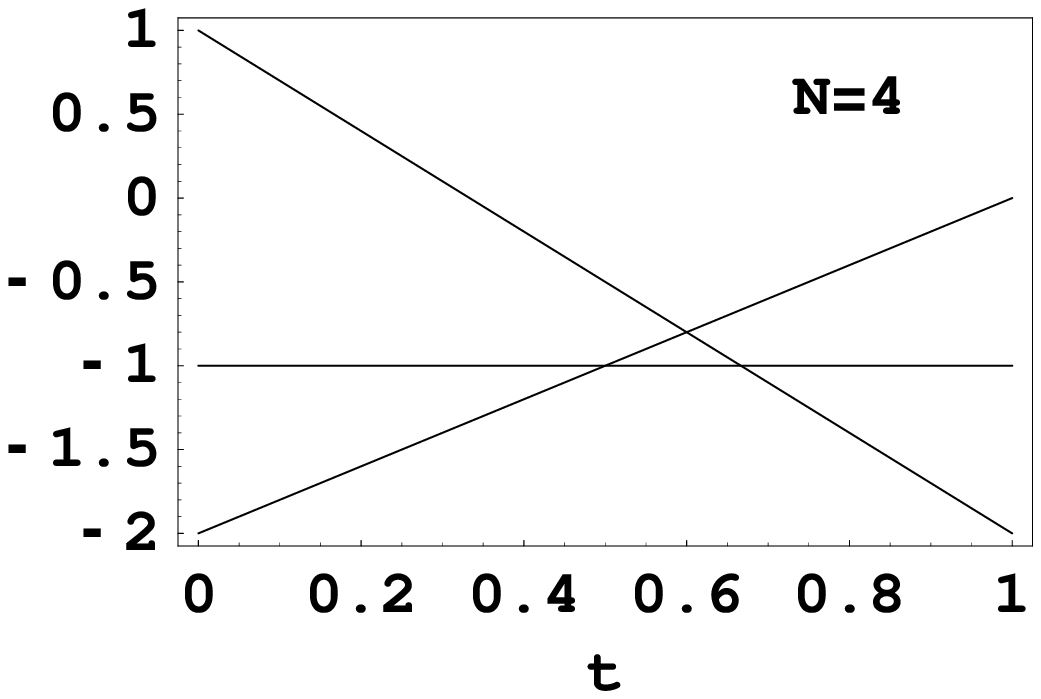,width=5cm}\epsfig{file=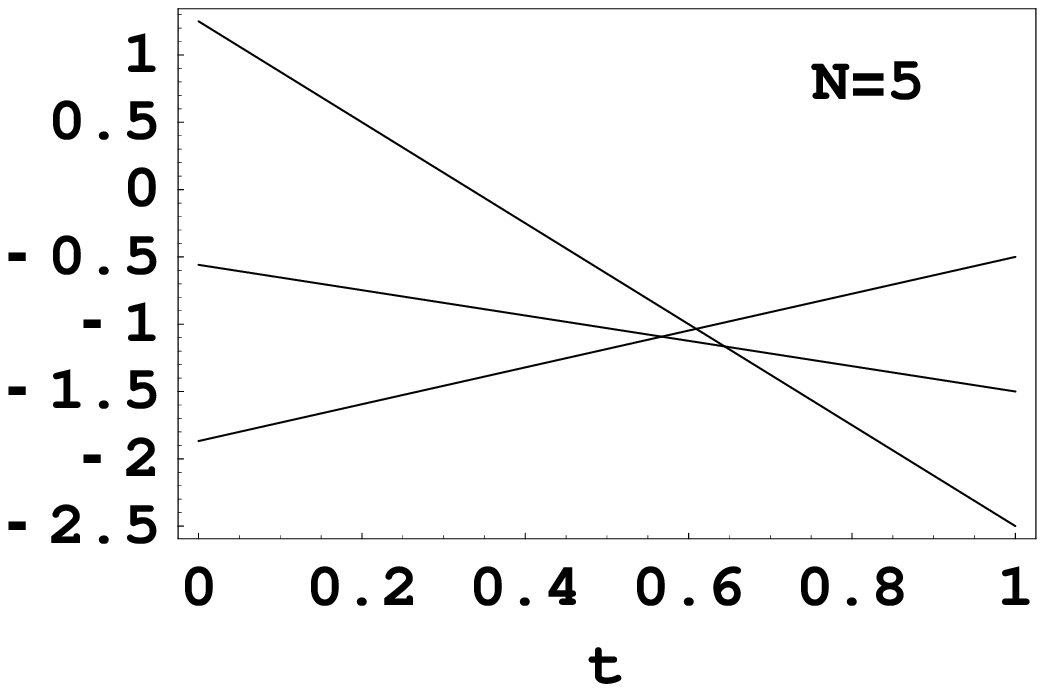,width=4.7
cm}}\\
\center{~~~~~\epsfig{file=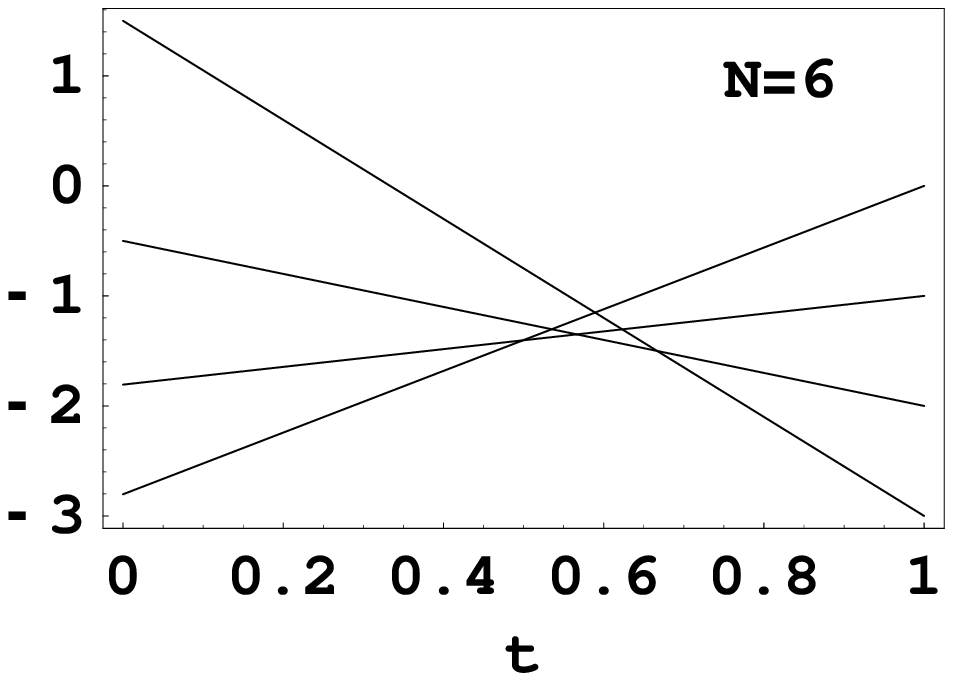,width=5cm}\epsfig{file=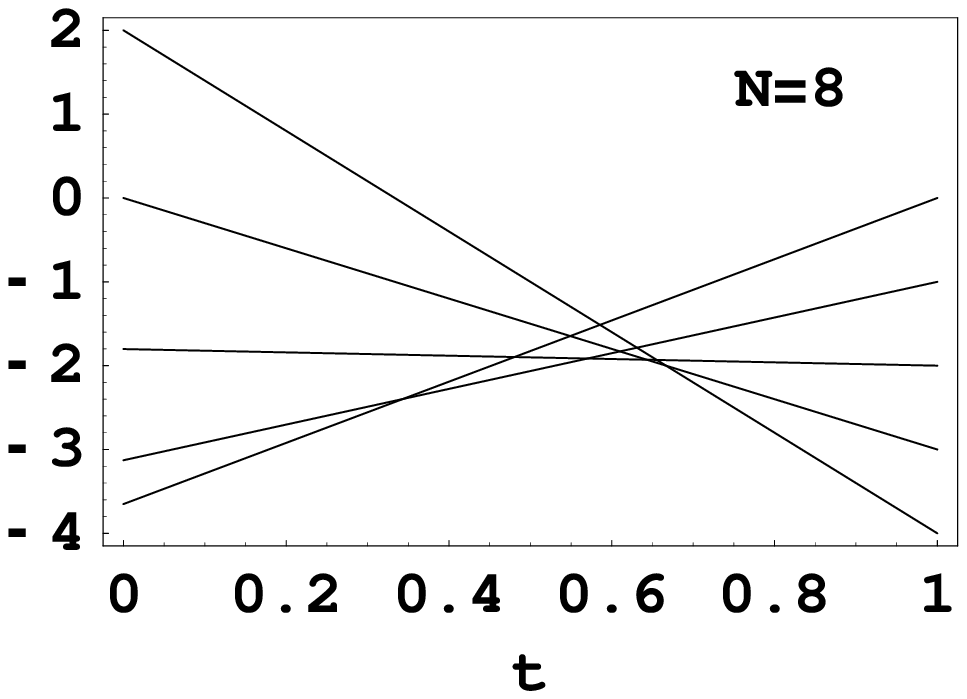,width=5cm
}\epsfig{file=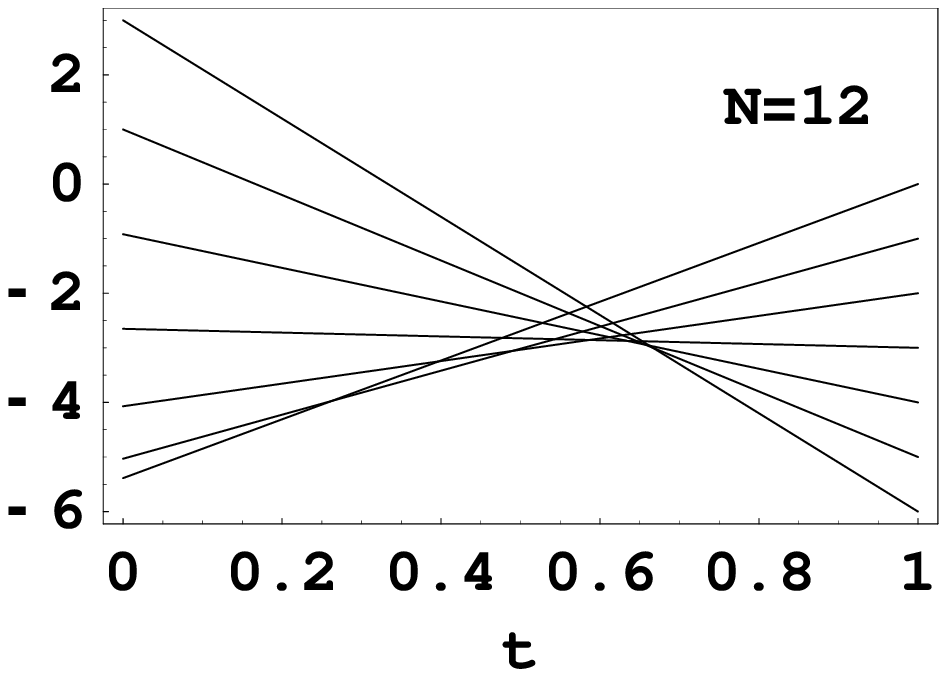,width=5cm }} \caption{Level crossings related
to the ground state of the anti-ferromagnetic case for different $N$
values.}
\end{figure}
\end{center}

   Entanglement measure in the model is one of important quantities
to characterize its QPT behavior, and is often studied by using
block--block entanglement defined in terms of von Neumann
entropy$^{[15]}$ or by using Wootters concurrence$^{[21]}$, e. g.,
that shown in [22]. In the following, we use the simple measure
proposed in [17-19] with

$$\eta(\Psi)=
-{1\over{N}}\sum^{N}_{i=1}{\rm Tr}\left\{
(\rho_{\Psi})_{i}\log(\rho_{\Psi})_{i} \right\}\eqno(6)$$ if all $N$
terms in the sum are non-zero, otherwise $\eta(\Psi)=0$, where
$\Psi$ stands for the ground state wavefunction and
$(\rho_{\Psi})_{i}$ is the reduced density matrix with $i$-th
spin-${1\over{2}}$ fermion only. It has been shown$^{[17-19, 23,
24]}$ that (6) is also suitable to measure genuine $N$-body
entanglement in a quantum many-body system. We observed that
$(\rho_{\Psi})_{i}$ is $i$-independent for the ground state in the
system for $N$ even cases, while it becomes $i$-dependent for $N$
odd cases. Hence, the entanglement measure $\eta$ for $N$ even cases
can be simply defined by the reduced von Neumann entropy for any
site, while it should be calculated separately for $N$ odd cases.
Table 2 shows ground state entanglement in different $t$ ranges for
$N=2,\cdots,6$, respectively, in which the entanglement type of the
ground state in each $t$ range is indicated. For example, the state
is a linear combination of several GHZ-like states for $N=4$ with
$0\leq t< 0.5$, while it consists of two-fold degenerate pair which
are all linear combinations of serval W-like states for $N=5$ with
$0\leq t<0.566915$. It is clear that the ground state entanglement
measure gradually increases while the control parameter $t$
decreases, which is also characterized by the quantum numbers $S$
and $S^{0}$. In the ferromagnetic (unentangled) phase, $S=N/2$ and
$S^{0}$ reaches its lowest value with $S^{0}=-N/2$, while
$S=S_{0}=0$ ($S=-S_{0}=1/2$) when $t<t_{c}^{(1)}$ for $N$ even
(odd), in which the spin-up and -down fermions are most strongly
correlated in comparison to that in other phases. In the most
entangled long-range order phase, $N$ even systems are most
entangled with $\eta=1$ which is always greater than those of the
nearest $N$ odd systems. Furthermore, the degeneracy is doubled at
the level-crossing points $t=t^{(j)}_{\rm c}$. For $N$ even cases,
the ground state is not degenerate if the control parameter $t$ does
not at those $[N/2]$ level-crossing points, while it becomes
two-fold degenerate when $t=t_{c}^{(j)}$ for any $j$ due to the
level-crossing. For $N$ odd cases, the ground state is four-fold
degenerate at $t=0$ and is a singlet when $t>t^{([N/2])}_{\rm c}$.
Besides those two cases, the ground state is two-fold degenerate
with $S_{0}=-S=-k+N/2$ for $k=0,1,2,\cdots,[N/2]$ if the control
parameter $t$ does not at those $[N/2]$ level-crossing points, while
it becomes four-fold degenerate when $t=t_{c}^{(j)}$ for any $j$ due
to the level-crossing. However, these degenerate states at the
level-crossing points are still distinguishable from each other by
the quantum number $S$ and $S^{0}$ with their difference
$\Delta(S^{0})=\Delta(S)=\pm 1$ and by values of the entanglement
measure of the degenerate states. As a consequence,  for $N$ even
case, the ground state is not degenerate when $t=0$; it becomes
two-fold degenerate everywhere when the control parameter $t$ is
within the half-open interval $t\in (0, 2/3]$ because the
level-crossing points are dense everywhere in this control parameter
range in the $N\rightarrow\infty$ limit; and finally it becomes not
degenerate again when $2/3<t\leq 1$. For $N$ odd case, the ground
state is four-fold degenerate for $t$ being within the closed
interval $t\in [0, 2/3]$ in the $N\rightarrow\infty$ limit; and it
becomes not degenerate when $2/3<t\leq 1$. Nevertheless, the
property of the degenerate states at $t=0$ and that within $0<t\leq
2/3$ are different for $N$ odd case. The four-fold degenerate states
at $t=0$ come from the double occurrence of $S=1/2$, while two
states from $S_{0}=-S=-k+N/2$ and another two from
$S_{0}=-S=-(k+1)+N/2$ to form the corresponding four-fold degeneracy
for $0<t\leq 2/3$. However, it has been proved at least for small
$N$ cases that GHZ- and W-type states are inequivalent under the
SLOCC transformations.$^{[23-25]}$ Therefore, the ground state
should be classified into three phases in the thermodynamic limit
for $N$ even case under the SLOCC. These three phases are one
non-degenerate entangled GHZ-type phase at $t=0$ with $\eta=1$, one
two-fold degenerate entangled W-type phase with $t\in (0,2/3]$ and
$0<\eta<1$, and one non-degenerate fully separable phase with $t\in
(2/3,1]$ and $\eta=0$. But such QPT classification is only
meaningful under the SLOCC. For $N$ odd case, the situation is
different. There is one four-fold degenerate entangled W-type phase
with $t\in [0,2/3]$ and $0<\eta<1$, and one non-degenerate fully
separable phase with $t\in (2/3,1]$ and $\eta=0$.

\vskip .4cm \noindent{\bf Table 2.}~{Ground state entanglement with
each quantum phase for $N=2,\cdots,6$.}\\{\small
\begin{tabular*}{\textwidth}{lllll}
\hline \hline $N$~~~~~~~~~~~~~~~~~~~~~~~~~~~~~~~
~~~~~~~Entanglement type in each phase\\
\hline\\
2~~{\begin{tabular}{c} $S=-S_{0}=1$\\
Fully separable\\
($\eta=0$)~$2/3< t\leq 1$\\
\end{tabular}}
~{\begin{tabular}{c}
$S=-S_{0}=0$\\
Bell~($\eta=1$)\\
 $0\leq t<2/3$\\
\end{tabular}}\\\\
3~~{\begin{tabular}{c}
$S=-S_{0}=3/2$\\
Fully separable\\
($\eta=0$)~$0.6< t\leq 1$
\end{tabular}}
~{\begin{tabular}{c}
$S=-S_{0}=1/2$\\
$\vert\xi=1\rangle$ is partially separable\\
($\eta_{1}=0$)\\
$\vert\xi=2\rangle$ is a W Combination\\
($\eta_{2}=0.739447$)\\
 $0\leq t<0.6$\\
\end{tabular}}\\\\
4~~{\begin{tabular}{c}
$S=-S_{0}=2$\\
Fully separable\\
($\eta=0$)~$2/3< t\leq 1$
\end{tabular}}
~{\begin{tabular}{c}
$S=-S_{0}=1$\\
W ~($\eta=0.811278$)\\
 $0.5<t<2/3$\\
\end{tabular}}
~{\begin{tabular}{c}
$S=-S_{0}=0$\\
GHZ Combination\\
($\eta=1$)\\
 $0\leq t<0.5$\\
\end{tabular}}\\\\
5~~{\begin{tabular}{c}
$S=-S_{0}=5/2$\\
Fully separable\\
($\eta=0$)\\
$0.644004< t\leq 1$
\end{tabular}}
~{\begin{tabular}{c}
$S=-S_{0}=3/2$\\
$\vert\xi=1\rangle$ is a W Combination\\
($\eta_{1}=0.610281$)\\
$\vert\xi=2\rangle$ is a W  combination\\
($\eta_{2}=0.619557$)\\
 $0.566915<t<0.644004 $\\
\end{tabular}}
~{\begin{tabular}{c}
$S=-S_{0}=1/2$\\
$\vert\xi=1\rangle$ is a W combination\\
($\eta_{1}=0.858927$)\\
$\vert\xi=2\rangle$ is a W combination\\
($\eta_{2}=0.858501$)\\
 $0\leq t<0.566915$\\
\end{tabular}}\\\\
6~~{\begin{tabular}{c}
$S=-S_{0}=3$\\
Fully separable\\
($\eta=0$)~$2/3< t\leq 1$\\
\end{tabular}}
~{\begin{tabular}{c}
$S=-S_{0}=2$\\
W ($\eta=0.650022$)\\
 $0.566401<t<2/3$\\
\end{tabular}}
~{\begin{tabular}{c}
$S=-S_{0}=1$\\
W Combination\\
($\eta=0.918296$)\\
 $0.499123 < t<0.566401$\\
\end{tabular}}
~{\begin{tabular}{c}
$S=-S_{0}=0$\\
GHZ Combination\\
($\eta=1$)\\
 $0\leq t<0.499123$\\
\end{tabular}}\\\\
\hline \hline
\end{tabular*}}

\vskip .4cm In summary, a Mathematica code based on the differential
realization of hard-core boson operators for constructing energy
matrix of the periodic-$N$ spin-${1/2}$ systems with or beyond
nearest neighbor interactions is proposed, which can easily be used
to study general spin-${1/2}$ interaction systems, such as XY or XYZ
spin-${1/2}$ chains. As an example, The code is applied to study the
anti-ferromagnetic XXX spin-${1/2}$ chain with nearest neighbor
interaction in a uniform transverse field. The study shows how the
ground state of the model evolves from the ferromagnetic phase to
the anti-ferromagnetic long-range order phase with decreasing of the
control parameter $t$ introduced.  In addition, we have shown that
there are $[N/2]$ level-crossing points, of which the middle part
will become a continuous one in the large-$N$ limit leading to the
three-phase result in the thermodynamic limit for $N$ even case
under the SLOCC, while there are only one entangled W-type phase and
one separable phase in the large $N$ limit for $N$ odd case. Such
level-crossing should be common in other spin interaction systems in
a uniform transverse field.

\vskip .5cm  Support from the U.S. National Science Foundation
(0500291), the Southeastern Universities Research Association, the
Natural Science Foundation of China (10575047), and the LSU--LNNU
joint research program (C192135) is acknowledged.

\end{document}